# Superballistic flow of viscous electron fluid induced by microwave irradiation in quantum point contact

Xinghao Wang[*], Wenfeng Zhang and Rui-Rui Du

International Center for Quantum Materials, School of Physics, Peking University, Beijing 100871, China

L. N. Pfeiffer, K. W. Baldwin, and K. W. West

Department of Electrical Engineering, Princeton University, Princeton, NJ 08544, USA

## Abstract

We measure the resistance oscillation of quantum point contact (QPC) under microwave (MW) radiation. What is different from the common resistance oscillation induced by edge magnetoplasmon (EMP) is that at lower magnetic field ($\omega > \omega_c$), photoconductance is positive (negative) with weak (strong) MW radiation. This transport phenomenon is proved to be related to superballistic flow of electrons through QPC. The distinction between the regions $\omega > \omega_c$ and $\omega < \omega_c$ is attributed to different absorption rate of MW radiation. Violent absorption occurs when cyclotron orbits or current domains are destroyed in QPC region.

[*] 17wxhwypku@pku.edu.cn

*Introduction.* - Hydrodynamic charge transport in high-mobility two dimensional electron gas (2DEG) has aroused much attention since viscosity of electron was verified in electron systems, e.g., graphene [1,2] and GaAs/AlGaAs quantum well (QW) [3-9]. Hydrodynamic regime induced by collective motion of electrons is reached when momentum conserving collisions between electrons dominate, which corresponds to the condition that mean free path (MFP) of electron-electron collision $l_{ee}$ is much shorter than both sample width $W$ and MFP of electron-impurity/phonon collision $l_0$. There are some prominent phenomena relevant to hydrodynamic charge transport, e.g., superballistic flow through narrow constrictions [2], negative resistance resulting from wirlpools of electron flow [8] and negative magnetoresistance (NMR) of Poiseuille flow [9,10]. Superballistic transport is the most interesting and counterintuitive phenomenon among them [2]. Narrow constriction in 2DEG forms a quantum point contact (QPC) and it has a quantum conductance decided by channel number $N_{QPC}$ multiply $2e^2/h$ for ballistic transport [11,12]. Interestingly, collective movement of viscous electrons can shield carriers from momentum loss at sample edges, thus resulting in larger QPC conductance.

In our previous work [9], hydrodynamic charge transport under microwave (MW) radiation in ultrahigh-mobility GaAs/AlGaAs 2DEG is studied through size dependence of second harmonic peak [10] and microwave-induced resistance oscillation (MIRO) [13-16]. For wide samples, i.e., $W \gg v_F\tau_2$ ($v_F$ is Fermi velocity and $\tau_2$ is the relaxation time of the second moment of electron distribution function), theories [17-22] predicted that MIRO amplitude is linear with $W$. For narrow samples, i.e., $W \ll v_F\tau_2$, MIRO is too weak to be detected. Naively thinking, there should be nothing interesting when sample width decreases to microns or submicron.

Surprisingly, we find out that there is another kind of resistance oscillation for QPCs under MW radiation. It is induced by interference of edge magnetoplasmon (EMP) [23-27] and severely relies on MW power $P_{MW}$. Under strong MW radiation, resistance oscillation shows huge and violent resistance peaks. While under weak MW radiation, resistance maximum astonishingly turns into resistance minimum, which is a sign of superballistic transport. Considering size dependence of radiation heating effect [9], we eventually reach a conclusion that QPC is extremely hot when strong MW absorption happens, with electron temperature $T_e$ even around $100\,K$ while ambient temperature $T_b$ is about $2K$.

The reason for the great MW absorption in QPC is very intriguing and is still an open question. It seems to be related to MIRO or zero-resistance state (ZRS). For high mobility 2DEG, some low-order minima of MIRO cannot form negative resistance but transform to ZRS [15,16,28], because homogeneous state with negative resistivity is electrically unstable and naturally breaks into current domains [29-32]. We find that resistance minima/maxima of the novel resistance oscillation appear only when bulk 2DEG is in the regime of MW induced ZRS. This is not a coincidence because narrow constriction like QPC can break down current domain of ZRS.

*Experimental setup.* - Our experiment is performed in a ³He refrigerator (base temperature $T_b = 0.3K$) with two wafers. After being illuminated by red light-emitting diode at $2K$, one is high-mobility 2DEG with $\mu$ around $2 \times 10^6 cm^2/Vs$ and the other has better quality with $\mu$ above $2 \times 10^7 cm^2/Vs$ at $0.3K$. Carrier density $n$ is both around $2.6 \times 10^{11}\,cm^{-2}$. Each sample consists of several QPCs with different constriction widths $W$ from $3.2\mu m$ to $0.2\mu m$. We use two kinds of QPCs. One is normal split gate QPC defined by e-beam lithography and Ti/Au top gate, and the other is constricted by wet etching without metal gate. Electrical contacts are made by Ge/Pd/Au alloy annealed at $450°C$. For details of each sample, please refer to Table SI of [33]. In our experiment, MW with frequency $f$ ranging from $30GHz$ to $105GHz$ is generated by Gunn oscillators and MW power $P_{MW}$ is adjusted by a programmable rotary vane attenuator. Resistance is measured with low-frequency lock-in technique.

We measure the longitudinal resistance $R_L$ through each QPC for the sake of symmetry of magnetoresistance (Fig.1(a)). According to Landauer-Buttiker formula [34],

$$R_L = \frac{h}{2e^2}\frac{1}{N_{QPC}} - |R_{xy}|, \quad (1)$$

Since Hall resistance $R_{xy}$ is not affected by MW radiation, the measured response of $R_L$ to MW fully reflects that of QPC resistance $R_{QPC}$. Effective width $W$ of QPC is decided by Sharvin formula,

$$R_b = \frac{h}{2e^2} \frac{\sqrt{\pi}}{W\sqrt{2n}}, \qquad (2)$$

where the ballistic resistance $R_b = h/(2e^2 N_{QPC})$ is the measured value of $R_{QPC}$ in zero magnetic field without MW. Under MW radiation, $R_{QPC}$ deviates from $R_b$ and we can measure its value in the following method. Since MW has little influence on $R_{xy}$, we obtain that $R_{QPC,MW} = R_{L,MW} + |R_{xy}| = R_{L,MW} - R_{L,0} + R_b$. Here, $R_{QPC,MW}$, $R_{L,MW}$ and $R_{L,0}$ denotes the resistance with or without MW radiation respectively.

Considering the fact that $R_L$ is mixed with $R_{xx}$ in the measurement, we note that $R_{L,MW}$ also incorporates MIRO signal in the bulk 2DEG. Due to high mobility of our samples, the mixed $R_{xx}$ is only several Ohms and does not influence the measured value of $R_{QPC}$.

*Characteristics of the novel resistance oscillation.* - Fig.1(b) shows overall feature of the novel resistance oscillation. There are two regions split apart by cyclotron resonance $\omega = 2\pi f = \omega_c$, where $\omega_c = eB/m^*$, $m^* = 0.067 m_e$ is effective mass of electron in GaAs QW, $B$ is magnetic field. Before introducing the novel resistance oscillation, we first notice that B-periodic oscillation in the high field region ($\omega < \omega_c$) is unambiguously attributed to EMP [23-27]. EMPs excited by MW propagate along the sample edge and interfere with each other in the QPC. The oscillation period of EMP satisfies $\Delta B \propto 1/\omega$, which is verified in our samples [23].

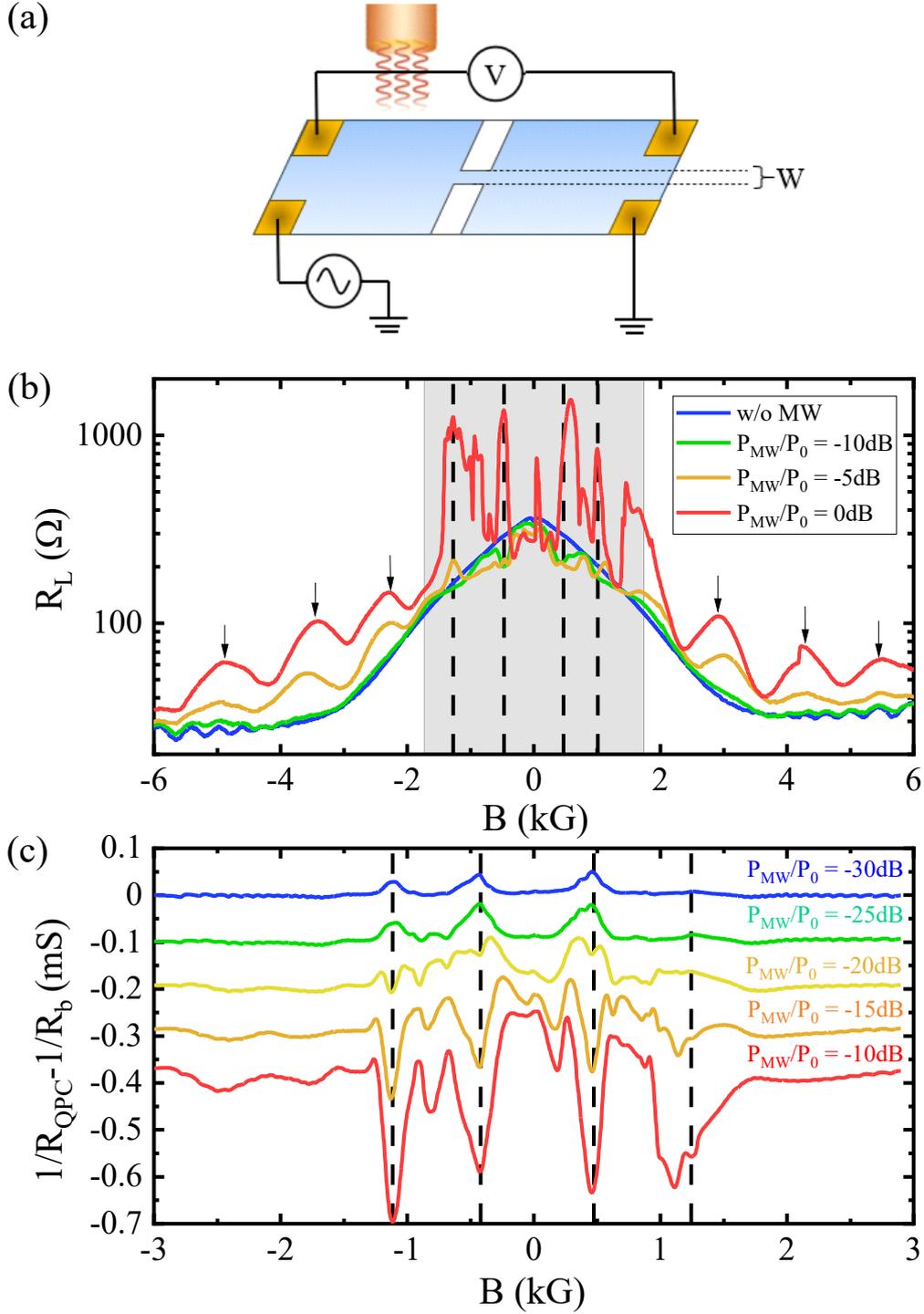

FIG. 1. (a) General setup of resistance measurement: a QPC with width $W$ under MW irradiation can bring about peculiar photoconductance. (b) The longitudinal resistance shows EMP-induced resistance oscillation and there is a crossover at $\omega = \omega_c$. In high magnetic field, the oscillation is $B$-periodic (arrows), while at $\omega > \omega_c$ (grey background) positive-to-negative photoconductance pattern and strange behavior of oscillations appear. Dash lines indicate positions of resistance minima/maxima. The attenuation $P_{MW}/P_0$ ranges from $-10dB$ to $0dB$. Background temperature $T_b$ is kept at $3K$. MW frequency $f$ is $71GHz$. This result is from QPC 1b, which is a split gate one with effective width $W$ around $0.9\mu m$ and gate voltage $V_{QPC} = -2.2V$. (c) Positive-to-negative photoconductance transformation (dash lines) happens when $P_{MW}/P_0$ is set from $-30dB$ to $-10dB$. Photoconductance for different MW power is consecutively shifted downward by $0.1mS$. The data is from QPC 2d, an etched one with effective width $W$ around $0.2\mu m$ at $T_b = 0.3K$.

More interesting region is in low magnetic field ($\omega > \omega_c$), where periodicity of the resistance oscillation seems elusive. For the example in Fig.1(b), when $P_{MW}$ is $-10dB$ multiplying power of Gunn oscillator $P_0$, photo-induced resistance $R_{L,MW} - R_{L,0}$ is negative at some magnetic field (positive photoconductance). When $P_{MW}/P_0 = 0dB$, dips of $R_{QPC}$ magically develop into huge peaks whose resistance is much larger than those at $\omega < \omega_c$ and multiple times of $R_L$ without MW (negative photoconductance). Similar phenomena appear in both split gate and etched QPCs, both high- and ultrahigh-mobility samples as well. For clarity, photoconductance $1/R_{QPC,MW} - 1/R_b$ is calculated for the data of another QPC (Fig.1(c)). The positive-to-negative photoconductance transformation is main feature of $\omega > \omega_c$ region.

In our previous work [23], we studied the EMP-induced resistance oscillation in relatively low-mobility samples and we did not focus on the positive-to-negative photoconductance transformation because it is more obvious for ultrahigh-mobility samples. According to paper [23], the peculiar resistance oscillation in our ultrahigh-mobility samples also comes from interference of EMPs. At interference construction point of EMPs, electrons sufficiently absorb MW irradiation, which brings about huge photoconductance. While at interference destruction point, QPC resistance $R_L$ is hardly influenced by MW. The data of ultrahigh-mobility sample is more interesting because of good quality of 2DEG and important role played by hydrodynamic transport.

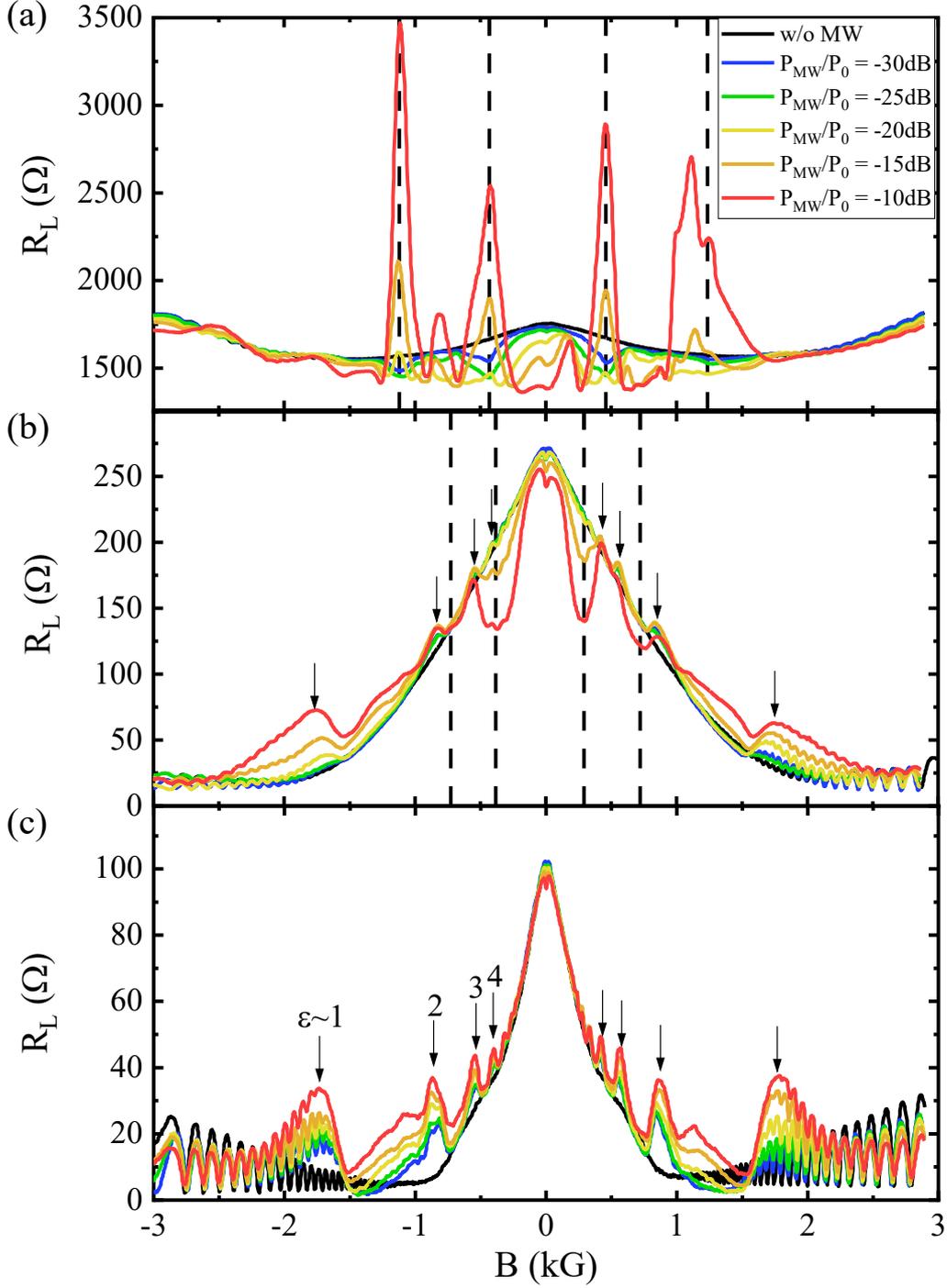

FIG. 2. The EMP-induced resistance oscillation in different QPCs under MW radiation with variant power together with bulk MIRO signal is shown: (a) QPC 2d, $W = 0.2\mu m$, (b) QPC 1a, $W = 1.2\mu m$, (c) QPC 2a, $W = 3.2\mu m$. Attenuation $P_{MW}/P_0$ ranges from $-30dB$ to $-10dB$. Arrows indicate observable low-order maxima of MIRO and dash lines mark the positions of resistance minima/maxima induced by EMPs. $T_b$ is kept at $0.3K$ and MW frequency $f$ remains $71GHz$.

Size dependence of the MW-induced resistance oscillation in QPC is demonstrated in Fig.2. For the narrowest QPC with $W = 0.2\mu m$ (Fig.2(a)), even very weak MW ($P_{MW}/P_0 < -30dB$) can result in resistance minima and they develop into maxima at $P_{MW}/P_0 = -15dB$. For QPC 1a with $W = 1.2\mu m$ (Fig.2(b)), MIRO appears in the background of NMR and the resistance minima do not turn into peaks until $P_{MW}/P_0 > -10dB$. Compared with QPC 2d, photoconductance $1/R_{QPC,MW} - 1/R_b$ of resistance minima in this wide QPC enhances dramatically. Photoconductance of QPC 2d is of the order of $0.1mS$ while that of QPC 1a is around $1mS$. For the widest constriction (Fig.2(c)), EMP-induced resistance oscillation doesn't appear but MIRO does. This indicates that the EMP-induced resistance oscillation only occurs in narrow QPCs. We arrive at a conclusion that wider QPC needs stronger MW to show the oscillation; but once there is EMP-induced resistance decrease, its photoconductance is much larger than the narrower one.

*Superballistic model.* - Strange pattern of the resistance oscillation at $\omega > \omega_c$ is hard to explain unless viscosity of electrons is considered. We notice that the positive-to-negative pattern of resistance variance frequently occurs in hydrodynamic electron systems, where electron-electron interaction plays an important role. An unusual decrease of QPC resistance with temperature was reported in previous work [35] and it was explained as electron-electron interactions mediated by boundaries. More recently, superballistic flow of viscous charge is studied in graphene constrictions [2]. QPC resistance first decreases and then increases as temperature goes higher. The former is attributed to interacting electrons shielding the momentum loss caused by boundaries and the latter is due to decrease of $l_0$. Viscosity of electrons modifies quantum conductance $G_b = 1/R_b$ to be $G_b + G_v$ and the viscosity contribution

$$G_v = \frac{\sqrt{2\pi n}e^2 W^2}{16\hbar v_F \tau_2}, \tag{3}$$

where $\hbar$ is Planck constant divided by $2\pi$. $G_v$ is calculated for the Stokes flow through a QPC in both hydrodynamic and ballistic regime [36]. This description is valid when phonon scattering does not dominate.

It is reasonable to think that the positive-to-negative photoconductance transformation is related to superballistic effect, for our samples are 2DEG of great quality where viscous electrons form collective flow at certain temperature. MW, absorbed by electrons passing through QPC, can significantly heat up the carriers in narrow constrictions, while electrons in the bulk remains cold. It makes sense since under strong MW radiation, $T_e$ of electrons in narrow constrictions is much higher than that in the bulk, and the latter is much higher than $T_b$ [9]. We are going to demonstrate that resistance of the minima/maxima in our experiment depends on MW power in the same pattern as resistance of superballistic electron flow depends on temperature.

Size dependence of the MW power sensitivity demonstrated in Fig.2 can be explained by size dependence of radiation heating effect [9]. Because of its lower conductance, narrower QPC absorbs more MW and its electrons are hotter. As a result, narrower QPC exhibits superballistic effect with weaker MW. Although narrow QPC response to MW sensitively, its photoconductance variance in hydrodynamic regime is small.

Size dependence of the photoconductance increase follows the rule of hydrodynamic transport. Maximum photoconductance of resistance minima $G_{v,max} = \left(1/R_{L,MW} - 1/R_{L,0}\right)_{max}$ under different MW power $P_{MW}$ is approximately proportional to $W^2$ (Fig.3(a)), which is exactly the same as superballistic transport in QPCs. The $W^2$ size dependence of photoconductance originates from viscous fluid self-organizing into streams with variant velocities in Poiseuille regimes.

Since viscous relaxation time $\tau_2$ can be extracted from the full width half maximum of NMR in 2DEG [17], we fit $1/\tau_2$ with $T_e$ with the following equation [21],

$$\frac{1}{\tau_2} \propto \frac{T_e^2}{\ln^2(\epsilon_F/k_B T_e)}, \qquad (4)$$

where $\epsilon_F$ is Fermi energy and $k_B$ is Boltzmann constant. Combining Eq.(3) and (4), we can then extract $T_e$ from $R_{QPC}$ in each resistance dip, assuming that electrons in QPC are still in hydrodynamic regime. We plot the extracted $T_e$ of four QPCs in Fig.3(b). $T_e$ is much higher than $T_b = 0.3K$ and positively related to $P_{MW}$. When $T_e$ is larger than $15K$, phonon scattering begins to dominate and the temperature extracted is lower than its real value. Saturation of calculated $T_e$ happens at about $15K$ for all the four QPCs (and the actual temperature after phonon scattering considered is higher, around $20 \sim 30K$ according to Fig.3(c) [33]), implying the crossover between superballistic and phonon scattering regime for GaAs 2DEG. Furthermore, we can fix the absorbed MW power $P_{a,MW}$ of all the QPCs to be the same value by simple translation of $P_{MW}$ axis for each QPC. $T_e$ only relies on $P_{a,MW}$ and MW absorption rate $P_{a,MW}/P_{MW}$ is size dependent [9]. Fig.3(c) shows the uniform relation between $T_e$ and $P_{a,MW}$. Assuming that actual $T_e$ is power function of $P_{a,MW}$, we make a logarithmic fit of our low temperature data which is not affected by phonon scattering, and extrapolate $T_e$ to larger $P_{MW}$.

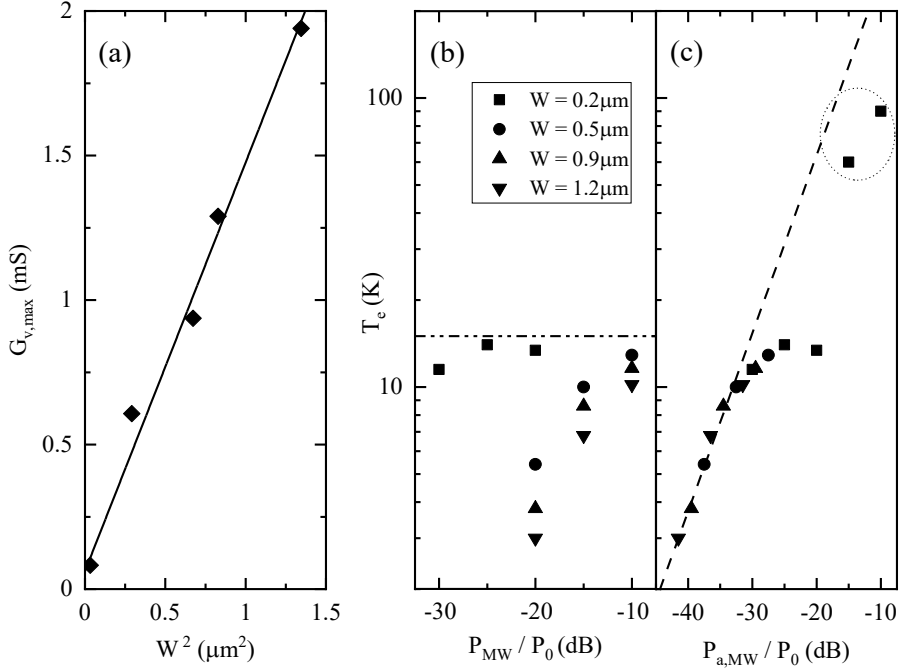

FIG. 3. (a) Maximum photoconductance of resistance minima $G_{v,max}$ is approximately proportional to $W^2$, which is a clear indicator of viscous charge transport. (b) The relation between $T_e$ at resistance minima and MW power is shown for four different QPCs (QPC 2d, $W = 0.2\mu m$; QPC 1b, $W = 0.5\mu m$; QPC 1b, $W =$

$0.9 \mu m$; QPC 1a, $W = 1.2 \mu m$). Since photo-induced resistance of a QPC may have several minima, we plot the average $T_e$ of two prominent symmetric minima of each trace. Saturation of $T_e$ due to phonon scattering is marked with dash-dot line. Subfigure (c) shows the result after $P_{a,MW}$ of all QPCs is changed to be the same as that of QPC 2d. Power law assumption of $T_e$ and $P_{MW}$ is shown as the dash line. $T_e$ estimated in diffusive regime qualitatively favors the trend of $T_e$ extracted in superballistic model.

In diffusive regime decided by phonon scattering, we can also estimate an approximate value of $T_e$. $R_{QPC}$ was calculated from ballistic to diffusive transport regime [37]. It is proved that $R_{QPC} = R_b + R_d$ and Drude resistance

$$R_d = \frac{h}{2e^2} \frac{2}{\sqrt{2\pi n} l_0} \frac{L}{W}, \qquad (5)$$

where $L$ is QPC length, which is about $1.2 \mu m$ for our samples. Given complicated scattering mechanism in GaAs QW, e. g. acoustic and optic phonon scatterings, we can roughly estimate the relation between $l_0$ and $T_e$ [38]. Then we obtain the estimated $T_e$ in diffusive regime from resistance peaks (Fig.3(c)). The estimated $T_e$ is lower than the extrapolated value in superballistic model, indicating clear enhancement of thermal conductance induced by electron-phonon coupling. When MW power is strong enough and huge resistance peaks appear ($P_{MW}/P_0 > 0dB$), mean free path $l_0 \ll L$ and $T_e$ could be even higher than $100K$. This fact interprets why EMP-induced resistance oscillation are so robust against high temperature.

*Discussion.* - Up till now, we have confirmed that the strange resistance oscillation at $\omega > \omega_c$ is induced by EMPs and, due to MW heating (just like micro-oven) at interference construction point of EMPs, superballistic effect result in positive photoconductance before negative photoconductance of EMPs appears. The strangest thing in our experiment is still unclear. That is the distinction between the region at $\omega > \omega_c$ and the one at $\omega < \omega_c$.

In paper [23], we attributed the distinction to different MW absorption rate. Photoconductance maxima appear when MW absorption rate reaches maxima and high $T_e$ results in superballistic/diffusive transport. When MW power is high, hot electrons passing through QPC inject into cold bulk 2DEG.

In a previous MW absorption experiment [39], only cyclotron resonance can make 2DEG absorb more MW, but what happens if it is a narrow constriction like QPC? The cyclotron radius is $r_c = 0.76 \mu m/B(kG)$, which is of the same order of QPC width $W$. We should notice in Fig.2 that the resistance minima or huge peaks (apart from cyclotron resonance) only appear when $r_c > W$. It seems that frequent collisions of electrons on the QPC boundary may magnify MW absorption rate and thus the EMP-induced resistance oscillation. This is because collisions of electrons can be related with photons. Lower conductance induced by collisions may also bring about high MW absorption rate.

Existence of MIRO and ZRS in the bulk 2DEG further complicates the problem. In our ultrahigh-mobility samples, high-quality MIRO and robust ZRS exist in the bulk connected by the QPC (Fig.4). Thus, it's interesting to discuss the interaction among bulk electrons and narrow constriction. When cyclotron orbits and current domains form in MIRO or ZRS and QPC width $W$ is smaller than their size, cyclotron orbits or current domains in QPC are completely destroyed. Electrons injected from source to drain break the equilibrium of MIRO/ZRS and result in negative resistance of MIRO [33]. Electrons transforming between MIRO/ZRS and QPC modes may absorb huge amount of MW. According to Fig.4, resistance minima/maxima of $R_L$ happen exactly when bulk electrons are in the regions of MIRO minima or ZRS. Once MIRO/ZRS is destroyed, electrons must consecutively absorb photons in order to maintain stability. This evidence unambiguously substantiates the relation between MW absorption and breaking of cyclotron orbits or current domains in the QPC region. Additionally, low conductance induced by MIRO minima/ZRS also contributes to high MW absorption rate.

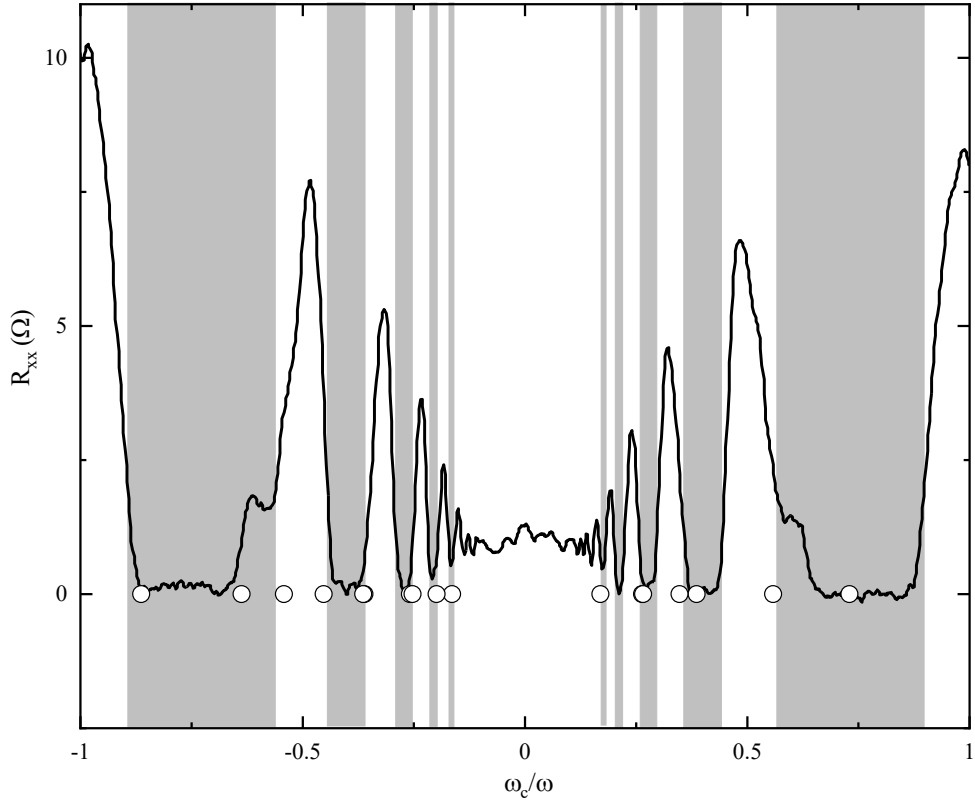

FIG. 4. MIRO in the bulk of ultrahigh-mobility 2DEG is demonstrated together with EMP-induced resistance minima/maxima marked by hollow circles and MIRO minima/ZRS are marked with grey background. We see that MW absorption rate maxima points tend to appear in the region of MIRO minima/ZRS. This is probably due to the fact that low conductance results in high MW absorption rate, especially when cyclotron orbits or current domains are destroyed in QPC region.

*Conclusion.* - We find out a novel EMP-induced resistance oscillation of QPC under MW radiation. The positive-to-negative photoconductance transformation is proved to be related to superballistic flow of electrons through QPC. The distinction between the regions $\omega > \omega_c$ and $\omega < \omega_c$ is attributed to different absorption rate of MW radiation. Violent absorption occurs when cyclotron orbits or current domains are destroyed in QPC region.

*Acknowledgements.* The work at PKU was funded by the National Key R&D Program of China (Grants No. 2017YFA0303300 and 2019YFA0308400), by the Strategic Priority Research Program of Chinese Academy of Sciences (Grant No. XDB28000000). The work at Princeton was funded by the Gordon and Betty Moore Foundation through the EPiQS initiative Grant No. GBMF4420, by the National Science Foundation MRSEC Grant No. DMR-1420541.

**References**

[1] D. A. Bandurin, I. Torre, R. K. Kumar, M. B. Shalom, A. Tomadin, A. Principi, G. H. Auton, E. Khestanova, K. S. Novoselov, I. V. Grigorieva, et al., "Negative local resistance caused by viscous electron backflow in graphene", Science **351**, 1055 (2016).

[2] R. K. Kumar, D. A. Bandurin, F. M. D. Pellegrino, Y. Cao, A. Principi, H. Guo, G. H. Auton, M. B. Shalom, L. A. Ponomarenko, G. Falkovich, et al., "Superballistic flow of viscous electron fluid through graphene constrictions", Nat. Phys. **13**, 1182 (2017).

[3] G. M. Gusev, A. D. Levin, E. V. Levinson, and A. K. Bakarov, "Viscous electron flow in mesoscopic two-dimensional electron gas", AIP Adv. **8**, 025318 (2018).

[4] A. D. Levin, G. M. Gusev, E. V. Levinson, Z. D. Kvon, and A. K. Bakarov, "Vorticity-induced negative nonlocal resistance in a viscous two-dimensional electron system", Phys. Rev. B **97**, 245308 (2018).

[5] G. M. Gusev, A. D. Levin, E. V. Levinson, and A. K. Bakarov, "Viscous transport and Hall viscosity in a two-dimensional electron system", Phys. Rev. B **98**, 161303(R) (2018).

[6] B. A. Braem, et.al., "Scanning gate microscopy in a viscous electron fluid", Phys. Rev. B **98**, 241304(R) (2018).

[7] A. C. Keser, et. al., "Geometric control of universal hydrodynamic flow in a two-dimensional electron fluid", Phys. Rev. X **11**, 031030 (2021).

[8] A. Gupta, J. J. Heremans, G. Kataria, M. Chandra, S. Fallahi, G. C. Gardner and M. J. Manfra, "Hydrodynamic and ballistic transport over large length scales in GaAs/AlGaAs", Phys. Rev. Lett. **126**, 076803 (2021).

[9] X. Wang, et. al., "Hydrodynamic charge transport in GaAs/AlGaAs ultrahigh-mobility two-dimensional electron gas", Phys. Rev. B **106**, L241302 (2022)

[10] Y. Dai, R. R. Du, L. N. Pfeiffer, and K. W. West, "Observation of a cyclotron harmonic spike in microwave-induced resistances in ultraclean GaAs/AlGaAs quantum wells", Phys. Rev. Lett. **105**, 246802 (2010).

[11] Y. V. Sharvin, "A possible method for studying Fermi surfaces", *Sov. Phys. JETP* **21,** 655–656 (1965).

[12] C. W. J. Beenakker & H. van Houten, "Quantum transport in semiconductor nanostructures", *Solid State Phys.* **44,** 1–228 (1991).

[13] M. A. Zudov, R. R. Du, J. A. Simmons, and J. L. Reno, "Shubnikov–de Haas-like oscillations in millimeterwave photoconductivity in a high-mobility two-dimensional electron gas", Phys. Rev. B **64**, 201311(R) (2001).

[14] P. D. Ye, L. W. Engel, D. C. Tsui, J. A. Simmons, J. R. Wendt, G. A. Vawter, and J. L. Reno, "Giant microwave photoresistance of two-dimensional electron gas", Appl. Phys. Lett. **79**, 2193 (2001).

[15] R. G. Mani, J. H. Smet, K. von Klitzing, V. Narayanamurti, W. B. Johnson, and V. Umansky, "Zero-resistance states induced by electromagnetic-wave excitation in GaAs/AlGaAs heterostructures", Nature **420**, 646 (2002).

[16] M. A. Zudov, R. R. Du, L. N. Pfeiffer, and K. W. West, "Evidence for a New Dissipationless Effect in 2D Electronic Transport", *Phys. Rev. Lett.* **90**, 046807 (2003).

[17] P. S. Alekseev, "Negative magnetoresistance in viscous flow of two-dimensional electrons", Phys. Rev. Lett. **117**, 166601 (2016).

[18] P. S. Alekseev, "Magnetic resonance in a high-frequency flow of a two-dimensional viscous electron fluid", Phys. Rev. B **98**, 165440 (2018).

[19] P. S. Alekseev and A. P. Alekseeva, "Transverse magnetosonic waves and viscoelastic resonance in a two-dimensional highly viscous electron fluid", Phys. Rev. Lett. **123**, 236801 (2019).

[20] P. S. Alekseev, "Magnetosonic waves in a two-dimensional electron Fermi liquid", Semiconductors **53**, 1367 (2019).

[21] P. S. Alekseev and A. P. Dmitriev, "Viscosity of two-dimensional electrons", Phys. Rev. B **102**, 241409(R) (2020).

[22] P. S. Alekseev and A. P. Alekseeva, "Microwave-induced resistance oscillations in highly viscous electron fluid", arXiv preprint arXiv:2105.01035v2.

[23] Xinghao Wang, Wenfeng Zhang, Rui-Rui Du, L. N. Pfeiffer, K. W. Baldwin, and K. W. West, "Aharonov-Borm oscillation and Microwave-induced edge-magnetoplasmon modes enhanced by quantum point contact", preprint (2024).

[24] I. V. Kukushkin, et. al., "New Type of B-Periodic Magneto-Oscillations in a Two-Dimensional Electron System Induced by Microwave Irradiation", Phys. Rev. Lett. **92**, 236803 (2004).

[25] I. V. Kukushkin, et. al., "Miniature quantum-well microwave spectrometer operating at liquid-nitrogen temperatures", Appl. Phys. Lett. **86**, 044101 (2005).

[26] K. Stone, et.al., "Photovoltaic oscillations due to edge-magnetoplasmon modes in a very high-mobility two-dimensional electron gas", Phys. Rev. B **76**, 153306 (2007).

[27] A. D. Levin, et. al., "Giant microwave-induced *B*-periodic magnetoresistance oscillations in a two-dimensional electron gas with a bridged-gate tunnel point contact", Phys. Rev. B **95**, 081408(R) (2017).

[28] R. L. Willett, L. N. Pfeiffer, and K. W. West, "Evidence for Current-Flow Anomalies in the Irradiated 2D Electron System at Small Magnetic Fields", *Phys. Rev. Lett.* **93**, 026804 (2004).

[29] A. V. Andreev, I. L. Aleiner, and A. J. Millis, "Dynamical Symmetry Breaking as the Origin of the Zero-dc-Resistance State in an ac-Driven System", *Phys. Rev. Lett.* **91**, 056803 (2003).

[30] A. Auerbach, I. Finkler, B. I. Halperin, and A. Yacoby, "Steady States of a Microwave-Irradiated Quantum-Hall Gas", *Phys. Rev. Lett.* **94**, 196801 (2005).

[31] Finkler, I. G., and B. I. Halperin, "Microwave-induced zero-resistance states are not necessarily static", *Phys. Rev. B* **79**, 085315 (2009).

[32] I. A. Dmitriev, A. D. Mirlin, D. G. Polyakov, M. A. Zudov, "Nonequilibrium phenomena in high Landau levels", *Rev. Mod. Phys.* **84**, 1709 (2012)

[33] Supplementary materials.

[34] M. Buttiker, "Four-terminal phase-coherent conductance", Phys. Rev. Lett. **57**(14):1761-1764, (1986).

[35] V. T. Renard, et. al., "Boundary-Mediated Electron-Electron Interactions in Quantum Point Contacts", Phys. Rev. Lett. **100**, 186801 (2008).

[36] H. Guo, et. al., "Higher-than-ballistic conduction of viscous electron flows", *Proc. Natl Acad. Sci. USA* **114,** 3068–3073 (2017).

[37] M. J. M. de Jong, "Transition from Sharvin to Drude resistance in high-mobility wires", *Phys. Rev. B* **49,** 7778–7781 (1994).

[38] L. Pheiffer, et. al., "The role of MBE in recent quantum Hall effect physics discoveries", *Physica E* **20**, 57–64 (2003).

[39] S. A. Studenikin, M. Potemski, A. Sachrajda, M. Hilke, L. N. Pfeiffer and K. W. West, "Microwave-induced resistance oscillations on a high-mobility two-dimensional electron gas: Exact waveform, absorption/reflection and temperature damping", *Phys. Rev. B* 71, 245313 (2005).